# A Tutorial For Creating and Publishing Open Source Lisp Software

Robert Smith
Secure Outcomes Inc.
2902 Evergreen Pkwy Ste 200
Evergreen, CO 80439 USA
rsmith@secureoutcomes.net

## ABSTRACT
The proliferation and accessability of the Internet have made it simple to view, download, and publish source code. This paper gives a short tutorial on how to create a new Common Lisp project and publish it.

## Categories and Subject Descriptors
A.m [**General Literature**]: Miscellaneous; D.2.6 [**Software**]: Programming Environments; D.2.7 [**Software**]: Distribution, Maintenance, and Enhancement; D.2.13 [**Software**]: Reusable Software—*Reusable libraries*

## General Terms
Languages

## 1. PREREQUISITES
This tutorial assumes the reader has installed an implementation of Lisp as well as Zach Beane's Quicklisp package management system. A tutorial on how to install this is provided on Quicklisp's website[1].

We also assume that the reader has installed the Mercurial distributed source control management system.

This tutorial also assumes you have a dedicated directory for Lisp projects. We use `~/Source/My/` in this tutorial.

It is useful to tell Quicklisp about this directory by adding the following to your Lisp's initialization file[2]:

```
#+quicklisp
(progn
  (pushnew #P"~/Source/My/"
           ql:*local-project-directories*))
```

This allows local Quicklisp projects to be loaded from the stated directory.

Henceforth, we shall adopt the convention that lines beginning with '`$`' denote input to a shell (such as bash), and lines beginning with '`>`' denote input to a Lisp REPL.

## 2. BITBUCKET AND REPOSITORIES
Bitbucket, by Atlassian, is a free, online service for hosting public open source projects and personal private projects. It is the particular service used in this tutorial. Other popular services exist, such as github and gitorious.

### 2.1 Create an Account
To create a Bitbucket account, simply go to `www.bitbucket.org` and click "sign up". It will ask for some fundamental information, such as a username. When you have signed up and confirmed your account, log in by going to the same website, and clicking "log in" and entering your username and password information.

Once you are logged in, you can administer your account, and do things such as adding SSH keys.

Now we are able to create new repositories.

### 2.2 Create a Repository
To create a repository, click "Repositories" and "Create Repository". Type in the name of the new repository[3]. For this tutorial, we will use `cl-averages`—a library for mathematical averages. Next, *uncheck* the box that states "this will be a private repository". Lastly, make sure "Mercurial" is chosen as the repository type, and "Common Lisp" is selected for the language option. The other fields are optional. Click "Create repository", and a new online repository will have been created.

Now, open a command line terminal, and change to the directory in which you'd like to write the source code. On the website, you'll see "Clone this repository" with HTTPS and

---

[1] http://www.quicklisp.org/beta/
[2] For example, `~/.sbclrc` for SBCL.
[3] It is conventional to choose the same name as the name of the ASDF system, i.e., some lisp symbol like `my-project`.

Figure 1: Terminal interaction when cloning a repository.
```
$ cd ~/Source/My/
$ hg clone https://tarballs_are_good@bitbucket.org/tarballs_are_good/cl-averages
destination directory: cl-averages
no changes found
updating to branch default
0 files updated, 0 files merged, 0 files removed, 0 files unresolved
$ cd cl-averages/
```

SSH options. Click HTTPS and copy the clone command that appears just below into the terminal. See figure 1 for the sample terminal interaction. Now, our repository has been created on disk and we can change to the directory containing it.

## 3. CREATING A NEW PROJECT
The *de facto* standard for specifying modules of Lisp code is the "Another System Definition Facility" or *ASDF*.

Start up your Lisp REPL with Quicklisp loaded, and issue the following command to load Zach Beane's Quickproject library:

```
> (ql:quickload :quickproject)
```

This will allow us to easily create template source files for a new project, letting us avoid tediously writing boilerplate. Once it is loaded, we create our new project using the function make-project. As its first argument, it takes the path to the repository. It also takes a few keyword arguments. One argument, :author, should contain your first and last name.

```
> (quickproject:make-project "~/Source/My/cl-averages/"
                             :author "Robert Smith")
```

Once this is done, four files will be created in the specified directory:

**README.txt** The README of the project.

**cl-averages.asd** The ASDF template file.

**cl-averages.lisp** Principal source file for the project.

**package.lisp** The package declaration and exports.

We will go through one-by-one to create our project.

### 3.1 The README File
The README should contain a description of the software project, and how to use it. Initially it'll contain a "stub", which should be deleted and replaced with relevant details. For example, we will write

```
CL-AVERAGES is a library for computing mathematical
means.
```

### 3.2 The ASDF File
The ASDF file contains information about how the system loads the project along with other meta-data. There is lots of information on how ASDF systems are specified, and luckily, Quickproject has done most of the important work for us.

Two important pieces of information need to be filled out: a short description of the project and the license under which you wish to place the project. Simply fill in the description as a string after the :description keyword, and then the name of the license after the :license keyword.

Depending on the license chosen, the relevant license should also be added to a new file in the directory called LICENSE or LICENSE.txt.

If the project gets more complicated in the future, more :file clauses can be added with the relevant files.

See figure 2 for the full program listing.

### 3.3 Writing the Source Code
Initially, we will have an in-package form followed by the following placeholder text:

```
;;; "cl-averages" goes here. Hacks and glory await!
```

We simply delete the placeholder text and write our program. For libraries, one usually writes a collection of data structures, functions, and macros. A convention to follow is that for internal—private—items, we specify the documentation as a comment above the defining form, and for external—public—items, we specify the documentation string in the defining form as normal.

For our example, we have three internal functions: sum, product, and assert-non-null. Following, we have our external functions: arithmetic-mean, geometric-mean, and harmonic-mean—three different kinds of mathematical means.

See figure 3 for the full program listing.

### 3.4 Exporting Symbols From the Package
Once we are at least somewhat done with our program or library, we can expose some symbols in the main package. As said, Quickproject has already created a package for us in package.lisp. To export symbols, we just add an :export form with our list of external symbols to the defpackage form:

```
(:export #:arithmetic-mean
         #:geometric-mean
         #:harmonic-mean)
```

Optionally, we can add a package nickname by adding a `:nicknames` clause to the `defpackage` form:

```
(:nicknames #:avg)
```

See figure 4 for the full code listing.

## 3.5 Loading and Testing the Software

If our project was created in a Quicklisp-aware directory, then we can load it using Quicklisp. Simply call `ql:quickload`:

```
> (ql:quickload :cl-averages)
```

If an error occurs, fix the source code and re-issue the command.

Once we have successfully loaded it, we can test with some REPL interactions. For example:

```
> (avg:arithmetic-mean 5.3 2.0 4.6)
3.9666665
> (avg:geometric-mean 5.3 2.0 4.6)
3.6533215
> (avg:harmonic-mean 5.3 2.0 4.6)
3.3110006
```

## 4. COMMITTING AND PUSHING THE CODE

Once we have made enough changes to the code to have a functional library, we can commit the code. Note that it is not necessary to have any sort of *complete* library; in fact, it is encouraged to *commit early* and *commit often*.

Before we commit our changes, we need to add new files to the repository. We just issue `hg add` as follows:

```
$ hg add cl-averages.asd cl-averages.lisp \
         package.lisp README.lisp
```

Now we can *commit* our changes, which registers the changes to the repository. We issue `hg commit` as follows:

```
$ hg commit -m "Initial commit."
```

Note that `"Initial commit."` is a *commit message*, which is a small note about the changes we've made. If we later add another kind of mean, we might commit again with the message `"Added the weighted harmonic mean."`.

Once we have committed, we can *push* our changes to Bitbucket, using the `hg push` command as follows:

```
$ hg push
```

It will be followed with a few messages, and it will ask for your password. Type your Bitbucket password (which will not be displayed), and press enter, and a few more messages will be displayed.

Now our code will be live online. To view it, go to the project's Bitbucket page. For example, https://bitbucket.org/tarballs is the link to author's tutorial project. The source code can then be viewed and downloaded online.

## 5. GETTING THE PROJECT IN QUICKLISP

Adding projects to Quicklisp allows them to be distributed to anyone with an Internet connection and the Quicklisp client. Adding new projects to Quicklisp is relatively fast and easy.

Quicklisp projects are tracked on github, a service similar to Bitbucket. Go to https://github.com/ and click "Signup" and then "Create a free account". Here you will fill out a form similar to the Bitbucket form.

Once you have created an account, sign in, and go to https://github.com/ and click "New Issue". You will be presented with a form, in which you type the request to add a project. It is advantageous to state four pieces of data:

1. a short description of the project,

2. the author of the project (you),

3. the license of the project, and

4. a link to the Bitbucket repository where the code can be obtained.

For `CL-AVERAGES`, we might have the title

```
Please add CL-AVERAGES
```

and the body

```
CL-AVERAGES is a library for computing mathematical
averages.

Author: Robert Smith
License: Public Domain
Code: https://bitbucket.org/tarballs_are_good/cl-averages
```

Once completed, click "Submit new issue" and it will be added to the queue. After some time, one of two things might happen: either the code will be accepted and added to Quicklisp (in which case the issue will be responded to with "added to Quicklisp"), or it will be rejected with some kind of reason (e.g., it would not compile or it is not portable).

If it has been accepted into Quicklisp, then you are done; your open source project will be available to everyone.

## 6. ACKNOWLEDGMENTS
Thanks to Zach Beane for creating Quicklisp and Quickproject—as well as maintaining the constant service associated with them—making the barrier-to-entry less of a barrier and more of an entry for newcomers, and relieving experienced users' past frustration with managing library dependencies.

Figure 2: Sample ASDF file for the CL-AVERAGES library (cl-averages.asd).

```
;;;; cl-averages.asd

(asdf:defsystem #:cl-averages
  :serial t
  :description "CL-AVERAGES is a library for mathematical means."
  :author "Robert Smith"
  :license "Public Domain"
  :components ((:file "package")
               (:file "cl-averages")))
```

Figure 3: Sample Lisp source code (cl-averages.lisp).

```
;;;; cl-averages.lisp

(in-package #:cl-averages)

;;; Compute the sum of the numbers in LIST.
(defun sum (list)
  (reduce #'+ list :initial-value 0))

;;; Compute the product of the numbers in LIST.
(defun product (list)
  (reduce #'* list :initial-value 1))

;;; Check that LIST isn't empty.
(defun assert-non-null (list)
  (assert (not (null list)) (list) "LIST must be a non-null list."))

(defun arithmetic-mean (&rest numbers)
  "Compute the arithmetic mean of the numbers NUMBERS."
  (assert-non-null numbers)
  (/ (sum numbers)
     (length numbers)))

(defun geometric-mean (&rest numbers)
  "Compute the geometric mean of the numbers NUMBERS."
  (assert-non-null numbers)
  (expt (product numbers)
        (/ (length numbers))))

(defun harmonic-mean (&rest numbers)
  "Compute the harmonic mean of the numbers NUMBERS."
  (assert-non-null numbers)
  (/ (length numbers)
     (sum (mapcar #'/ numbers))))
```

Figure 4: Sample package declaration code (package.lisp).

```
;;;; package.lisp

(defpackage #:cl-averages
  (:use #:cl)
  (:nicknames #:avg)
  (:export #:arithmetic-mean
           #:geometric-mean
           #:harmonic-mean))
```